\documentclass[sigconf,nonacm]{acmart}
\usepackage{multirow}
\usepackage{siunitx}
\usepackage{xspace}
\makeatletter
\g@addto@macro\UrlBreaks{\do\@}
\makeatother

\renewcommand\footnotetextcopyrightpermission[1]{}
\newcommand{\pkg}[1]{\path{#1}}

\newcommand{\npm}{\texttt{npm}\xspace}
\newcommand{\pnpm}{\texttt{pnpm}\xspace}
\newcommand{\yarn}{\texttt{Yarn}\xspace}
\newcommand{\cargo}{\texttt{cargo}\xspace}
\newcommand{\trivy}{\texttt{Trivy}\xspace}
\newcommand{\syft}{\texttt{Syft}\xspace}
\newcommand{\cdxgen}{\texttt{cdxgen}\xspace}
\newcommand{\js}{\texttt{JavaScript}\xspace}
\newcommand{\rust}{\texttt{Rust}\xspace}

\newcommand{\covTrivyJSr}{\SI{30.25}{\percent} }
\newcommand{\covSyftJSr}{\SI{42.75}{\percent} }
\newcommand{\covCdxgenJSr}{\SI{99.91}{\percent} }

\newcommand{\covTrivyJSd}{\SI{32.71}{\percent} }
\newcommand{\covSyftJSd}{\SI{43.78}{\percent} }
\newcommand{\covCdxgenJSd}{\SI{98.72}{\percent} }

\newcommand{\covSyftRr}{\SI{100}{\percent} }
\newcommand{\covTrivyRr}{\SI{83.02}{\percent} }
\newcommand{\covCdxgenRr}{\SI{100}{\percent} }

\newcommand{\covSyftRd}{\SI{100}{\percent} }
\newcommand{\covTrivyRd}{\SI{83.85}{\percent} }
\newcommand{\covCdxgenRd}{\SI{99.34}{\percent} }

\acmConference[SCORED '26]{ACM Conference on Software Supply Chain Offensive Research and Ecosystem Defenses}{6 October 2026}{Prague, Czech Republic} 
\acmYear{2026}
\copyrightyear{2026}
\acmISBN{}
\acmDOI{}

\title[Mind the Gap: How SBOM Specification Ambiguities Lead to Divergent Software Bills of Materials.
]{Mind the Gap: How SBOM Specification Ambiguities Lead to Divergent Software Bills of Materials. An Empirical Tool Study.}

\begin{document}


\begin{abstract}
Software Bill of Materials (SBOMs) will become mandatory starting in
December 2027 under the European Cyber Resilience Act
(CRA)~\cite{cra2024}. Although previous studies have highlighted
significant differences among SBOM generators, the reasons for these
discrepancies remain unknown, as does whether they stem from
implementation errors or deliberate design choices.

In this paper, we evaluate three widely used SBOM generators across more than 3,000
JavaScript and Rust projects, using a ground-truth baseline derived
from dependency lockfiles.

Our results show that these tools diverge in terms of both dependency
coverage and SBOM completeness. Importantly, most of these
discrepancies are systematic rather than accidental: they arise from
differing assumptions regarding dependency scope, naming, provenance,
and representation, while others reflect inconsistent support for
fields defined in SBOM specifications.

These findings demonstrate that many of the observed discrepancies
cannot simply be "fixed": they require clearer standardization. As SBOM
generation becomes a legal compliance requirement, the choice of tool
itself can influence the resulting SBOM, potentially becoming a source
of undetected non-compliance. We argue that future SBOM standards
should define canonical rules regarding dependency scope, provenance,
and representation to improve interoperability and compliance.
\end{abstract}




\author{Alan Prado}
\affiliation{%
  \institution{Univ. Rennes, Inria, CNRS, IRISA}
  \city{Rennes}
  \country{France}
}
\email{alan.prado@inria.fr}

\author{Olivier Zendra}
\affiliation{%
  \institution{Univ. Rennes, Inria, CNRS, IRISA}
  \city{Rennes}
  \country{France}
}
\email{olivier.zendra@inria.fr}

\author{Philippe Boinot}
\affiliation{%
  \institution{ANSSI}
  \city{Rennes}
  \country{France}
}

\author{Olivier Barais}
\affiliation{%
  \institution{Univ. Rennes, Inria, CNRS, IRISA}
  \city{Rennes}
  \country{France}
}
\email{Olivier.Barais@irisa.fr}


\maketitle

\section{Introduction}
\label{sec:introduction}

Software Bill of Materials (SBOMs) are structured inventories of software dependencies, standardized through formats such as SPDX and CycloneDX\cite{spdx,cyclonedx}. They are now widely used as part of software supply chain security practices. 
A typical usage is vulnerability management: organizations cross-reference their component inventory with public vulnerability databases to identify known threats. 
SBOMs thus improve supply chain transparency and help assess the risks associated with third-party dependencies~\cite{odonoghue2025softwarematerialssoftwaresupply}.

This growing importance has prompted regulatory action. In the United States, the National Telecommunications and Information Administration (NTIA) established minimum requirements for SBOM content in 2021~\cite{ntia2021minimum}. 
In Europe, the Cyber Resilience Act (CRA)~\cite{cra2024} goes further, mandating SBOMs for all software products placed on the EU market, with full compliance and enforcement due by December 2027.
Implicitly, both frameworks assume that SBOM generation tools accurately reflect the actual dependencies of a software product.
However, this assumption does not always hold. 

\medskip

In fact, several studies have evaluated SBOM generation tools and found that they often produce incomplete and inconsistent SBOMs, with different tools producing different results for the same project~\cite{Yu2024,Wang_2026,Tobar2025,rabbi2024sbom,rabbi2025claim}. Tools disagree on which dependencies to include, how to name them, and which versions to report. This has been observed across multiple languages and package managers. 
This can clearly undermine key use cases such as vulnerability detection~\cite{benedetti2024impactsbomgeneratorsvulnerability}.

These studies document the divergences and, in some cases, hint at possible causes, but none systematically explains their nature or origin at scale. 
How much of the divergence stems from normalization artifacts, deliberate design choices, or tool errors, and in what proportion, remains unclear. 
This is the gap our work aims to address.

\medskip 

In this paper, we study two ecosystems with different characteristics. 
\js has the largest package ecosystem in the world, built around a micro-package philosophy and spread across several package managers: \npm, \yarn, and \pnpm. 
\rust is a more recent language focused on memory safety and performance, with a single package manager, \cargo. Both ecosystems use lockfiles, files that pin every dependency (direct and transitive) to an exact resolved version, ensuring reproducible installs, which makes it possible to compare tool-generated SBOMs against a reliable reference.

We formulate three research questions:
\begin{description}
\item[RQ1] What is the coverage of \syft, \trivy and \cdxgen on \js and \rust projects, and how does it vary across dependency scopes?
\item[RQ2] What mechanisms generate divergences between tool output and the lockfile baseline, including on complex structures such as monorepos and multi-lockfile projects, and what is their nature: representation difference, design choice, or tool error?
\item[RQ3] How completely does each tool populate the fields a specification-compliant SBOM consumer would expect? 
\end{description}

\begin{sloppypar}
To answer our research questions, we built a dataset of 2,050 \js and 1,276 \rust projects from GitHub, archived via Software Heritage~\cite{swh}.
We generated SBOMs with three widely used tools, \syft~\cite{syft}, \trivy~\cite{trivy}, and \cdxgen~\cite{cdxgen}, and compared their output against our lockfile-based reference. 
We then applied a normalization pipeline to classify each divergence as a \textit{representation difference} (same dependency, different encoding), a \textit{design choice} (deliberate tool behavior), or a \textit{tool error}.
\end{sloppypar}

\smallskip
Our results show that \syft and \trivy cover only \covSyftJSd and \covTrivyJSd of \js dependencies, but reach \covSyftRd and \covTrivyRd on \rust. The tool \cdxgen performs well in both ecosystems, achieving \covCdxgenJSd to \covCdxgenJSr coverage in \js and \covCdxgenRd to \covCdxgenRr in \rust. When no lockfile is available, \trivy and \syft produce an empty SBOM. Our analysis also shows that most divergences are explainable, while some reflect understandable but incompatible design choices between tools. Beyond coverage, completeness for some specification fields also varies across tools. Together, these findings raise important questions for SBOM standardization, especially in the context of the CRA deadlines.

\smallskip

The rest of this paper is organized as follows. Section~\ref{sec:related-work} reviews related work. Section~\ref{sec:methodology} describes our dataset, reference, and normalization pipeline. Section~\ref{sec:results} presents our results. Section~\ref{sec:discussion} discusses our work and its limitations. Section~\ref{sec:conclusion} concludes and sketches future work.


\section{Background \& Related Work}
\label{sec:related-work}

Software Composition Analysis (SCA) tools have become a central
element of modern software supply chain security, 
by providing Software Bill of Materials (SBOMs), i.e., inventories of third-party components and their associated metadata written in standardized formats such as CycloneDX~\cite{cyclonedx} or SPDX~\cite{spdx}.
This enables enhancing software transparency and identification of third-party components, their dependencies, and associated vulnerabilities. 

\subsection{Context: Regulatory Drivers.}
SBOM adoption has accelerated in recent years due to growing concerns
regarding software supply chain security and the emergence of
regulatory requirements.

In the United States, government and executive initiatives have
encouraged SBOM adoption as a security practice for critical software
supply chains~\cite{eo14028}. 

Similarly, in Europe, regulatory developments are
reinforcing the need for transparent software composition
management. The Cyber Resilience Act (CRA)~\cite{cra2024}, in particular, introduces
cybersecurity obligations for products with digital elements,
including requirements regarding vulnerability management, secure
development practices, and software components transparency. Under
these obligations, manufacturers are required to maintain accurate
information on the software components integrated into their products,
thereby increasing the importance of reliable SBOM generation and
analysis.
\subsection{Tooling Ecosystem.}

However, the rapid shift from voluntary adoption to regulatory
requirements regarding SBOMs has led to the emergence of a vast
ecosystem of tools supporting software supply chain security
activities. Existing solutions cover various stages of the SBOM
lifecycle, including generation, usage, validation, transformation,
and interoperability. This analysis shows that SBOM generation remains the predominant use case, with 54 tools (64\%) offering generation capabilities, while standardization and interoperability remain persistent challenges within the ecosystem~\cite{mirakhorli2024landscapestudyopensource}.

Similarly, the "CycloneDX Tool Center" reports a growing number of
SBOM related tools with over 288 current
listings~\cite{cyclonedx-toolcenter}, thereby illustrating the rapid
expansion of the SBOM ecosystem.

\subsection{Limitations in Component and Dependency Extraction.}
While SBOM generation capabilities have significantly expanded,
compliance requires more than merely producing an SBOM document: the
generated inventory must be accurate, complete, interoperable, and
representative of the actual software composition. 
Consequently, weaknesses in dependency extraction, component identification, and
SBOM interpretation become critical challenges as organizations
increasingly rely on automated SCA pipelines to satisfy emerging
regulatory expectations.

However, recent studies have highlighted several limitations affecting
the reliability, interoperability, and completeness of current SCA
solutions.

\subsubsection{Component Identification.}
A first category of limitations concerns the accuracy of component
identification. 
Several studies~\cite{rabbi2024sbom,Wang_2026,Tobar2025} report that SCA tools frequently fail
to correctly extract component metadata, particularly package names
and versions, leading to incomplete or incorrect SBOM
entries. 

These
identification errors directly and adversely affect downstream vulnerability
analysis, since inaccurate component inventories may prevent the
detection of vulnerable dependencies or, conversely, generate incorrect security
alerts.

\subsubsection{Complex ecosystems.}
Another challenge relates to the analysis of complex package
ecosystems. Existing SCA approaches often rely on generic dependency
extraction mechanisms that struggle with ecosystems where dependency
resolution is managed through multiple package managers or
language-specific mechanisms. 
For instance, Python projects involving
\texttt{Poetry}, \texttt{Pipenv}, and \texttt{setuptools} introduce
additional metadata sources and resolution rules that are not always
correctly handled by SCA
tools~\cite{benedetti2024impactsbomgeneratorsvulnerability,cofano2024sbomgenerationtoolspython}. 
More generally,
several studies~\cite{Zhou_2026,benedetti2024impactsbomgeneratorsvulnerability} argue that multi-language SCA approaches should
incorporate the native dependency resolution mechanisms of each
programming language, instead of relying exclusively on universal
extraction strategies.

\subsubsection{Weak Dependency analysis.}
Dependency analysis itself remains a major source of
inaccuracies. Multiple studies identify problems in the reconstruction
of dependency graphs, including inconsistent dependency tree
roots~\cite{rabbi2024sbom} and incorrect exploration of transitive
dependencies across ecosystems such as \js, \rust, and
Python~\cite{rabbi2024sbom,Tobar2025,Yu2024,Halbritter2024,benedetti2024impactsbomgeneratorsvulnerability,kishimoto2025datasetsoftwarematerialsevaluating}.
Furthermore, SCA tools may 
report dependencies that are declared but unused, commonly referred 
to as bloated dependencies~\cite{rabbi2024sbom}. 
Conversely, they may include 
build-time, testing, or environment-specific dependencies, 
introducing false positives and contaminating the generated SBOM 
with components that are not part of the deployed 
application~\cite{rabbi2025claim,Tobar2025}.

\subsubsection{Syntactic Correctness and Parser Robustness.}
Beyond dependency extraction, related work highlights structural 
vulnerabilities regarding SBOM correctness. Specifically, current SCA 
tools may generate malformed documents or fail unexpectedly when 
processing syntactically valid SBOMs~\cite{Halbritter2024,kishimoto2025datasetsoftwarematerialsevaluating}. 
Such discrepancies in parser implementation not only hinder interoperability 
but also open the door to security risks, including parser confusion attacks~\cite{Yu2024}.




\section{Methodology} \label{sec:methodology}


Previous work shows that current Software Composition Analysis (SCA) solutions still face fundamental challenges in accurately identifying software components, reconstructing dependency relationships, producing interoperable SBOMs, and maintaining consistent interpretations across tools, often generating different SBOMs when analyzing the same source code. In this paper, we investigate the factors that cause different tools to produce distinct SBOMs from identical inputs. To better understand the origin of these discrepancies, we deliberately narrow the scope of our study by:
\begin{itemize}
\item considering only source code repositories;
\item excluding dynamic dependencies and runtime dependency resolution mechanisms.
\end{itemize}

Furthermore, we focus on projects that provide dependency manifests and, additionally, lockfiles (e.g., \texttt{yarn.lock}, \texttt{pnpm-lock.yaml}, and \texttt{Cargo.lock}). These artifacts allow us to establish a \textbf{ground-truth dependency oracle}, serving as a baseline for evaluating the accuracy of the generated SBOMs.

By controlling these variables, we aim to identify the factors that lead tools to report different dependency inventories despite analyzing identical source code.

This section now describes how we collected the dataset, generated SBOMs, extracted the lockfile baseline, and compared tool output against it.



\subsection{Dataset} \label{sec:dataset}

\subsubsection{Language Selection.}
We choose two different programming languages: \js and \rust.
\begin{itemize}
  \item \js has the largest and most interconnected package
    ecosystem in the world, largely due to its micro-package
    philosophy. It is also a prime target for dependency confusion
    attacks, typosquatting, and malicious package
    injection. Furthermore, the \js ecosystem is fragmented across several major package managers, including \npm, \yarn (Classic v1 and Berry v2+), and \pnpm.

  \item \rust can be considered the opposite of \js in this
    regard. It has gained considerable popularity thanks to its
    emphasis on memory safety, high performance, and a
    developer-friendly
    experience~\cite{bugden2022rustprogramminglanguagesafety}. Large
    technology companies are increasingly adopting \rust to build
    secure, robust, and scalable software systems. This growing
    adoption is reflected in the TIOBE Index, where \rust recently
    climbed to 12th place~\footnote{https://www.tiobe.com/tiobe-index/}.
\end{itemize}

Another reason for selecting these two languages is that both
ecosystems use dependency manifest files and lockfiles, making it
possible to analyze dependency declarations and the mechanisms used to
ensure reproducible dependency resolution.


\subsubsection{Project Selection.}
We queried the GitHub Search API to retrieve popular repositories in \rust and \js, sorted by star count in descending order, with a lower bound of 1,000 stars. We restrict our dataset to projects providing at least one manifest file (\texttt{package.json} for \js and \texttt{Cargo.toml} for \rust) declaring at least one dependency. Projects for which a fresh installation could not produce a lockfile are excluded from the dataset, as described in the dataset paragraphs below.

\subsubsection{Archiving and reproducibility.}
To ensure strict control over the projects under study, each selected project was matched against Software Heritage (SWH), a universal source code archive that stores repository snapshots as an immutable Merkle graph. Each revision is identified by a unique SWHID id (which is derived from the SHA-1 hash of the content tree)~\cite{swhid}, guaranteeing that two independent analyses of the same identifier will examine exactly the same repository state.
Only projects that had been archived by SWH at least once were retained. The revision corresponding to the most recent archival snapshot was selected.

\subsubsection{\js dataset.}

\begin{sloppypar}

We collected 2,105 \js projects. Rather than relying on lockfiles committed by developers, which may be desynchronized from the manifest or simply absent, we regenerated a fresh lockfile for every project in our dataset. To do so, we first had to determine which package manager to invoke: we inferred it from the existing lockfile when one was present in the repository (\npm, \yarn, or \pnpm), and otherwise consulted the packageManager field in \pkg{package.json}, falling back to \npm by default. This freezing step also allowed us to exclude projects that could not be installed. After this step, 2,050 projects had a successfully generated lockfile, failures being primarily due to yanked packages or unsatisfiable dependency constraints, and constitute our final dataset for \js.

\end{sloppypar}

\subsubsection{\rust dataset.}
 We collected 1,361 \rust projects. Rather than relying on \texttt{Cargo.lock} files committed by developers, which may be desynchronized from the manifest or simply absent, we performed a fresh \texttt{cargo fetch} for each project to generate a new \texttt{Cargo.lock}. This freezing step also allowed us to exclude projects for which the root installation failed. After this step, 1,276 projects had a successfully generated \texttt{Cargo.lock}, failures being primarily due to yanked packages or unsatisfiable dependency constraints, and constitute our final dataset for \rust.

\subsubsection{Lockfile-Based Reference}
\label{sec:lockfile-based-reference}

The goal is to establish a baseline against which the output of the SBOM generation tools can be compared. We cross-compare each tool's output against this baseline to identify which dependencies are missing from each tool's SBOM, and which are reported in addition to the baseline. Building this baseline requires parsing the lockfiles generated during the freezing step (Section~\ref{sec:dataset}), since they contain the exact, resolved set of dependencies produced by a fresh installation, independent of what developers originally committed.  We built parsers for all lockfile formats in our dataset: \npm \path{package-lock.json} in three versions (v1, v2, v3) plus \path{npm-shrinkwrap.json} (a v1-equivalent), \yarn v1 and Berry, \pnpm v5, v6 and v9 (v7 and v8 share the v6 format), and \path{Cargo.lock} v1/v2/v3.  Each package is annotated with its dependency type: production, development, optional, peer, or build (\rust).  Our parser extracts all dependencies present in the lockfile without filtering by type, to remain exhaustive. This allows us to compare and understand potential differences in tool behavior.

\subsubsection{Manifest-Only Setting}
\label{sec:manifest-only}

\begin{sloppypar}

To assess tool behavior when no pinned dependency tree is available, we additionally run SBOM generation tools on every project in the dataset after removing all lockfiles, leaving only the manifest   file (\texttt{package.json} or \texttt{Cargo.toml}). This setting is applied uniformly across the entire dataset, so that every project can be analyzed under both conditions, lockfile-based and manifest-only, and compared on a per-project basis. 
The reference baseline used for comparison differs by ecosystem. For \rust, \cdxgen analyzes \texttt{Cargo.toml} statically without performing any installation; the frozen \texttt{Cargo.lock}   generated during the freezing step (Section~\ref{sec:dataset}) thus serves directly as the reference, as no temporal drift can occur. For \js, \cdxgen performs a dynamic \texttt{npm install} at   analysis time; to ensure temporal comparability, we perform a concurrent fresh installation for each project and use the resulting lockfile as the reference baseline. This avoids biases introduced   by version drift between a pre-generated frozen lockfile and \cdxgen's runtime resolution.    We then compare the SBOM produced by each tool under the manifest-only setting against this reference to quantify the gap between the actual dependency tree and what each tool reports when only the   manifest is available.

\end{sloppypar}

\subsection{Experimental Setup}
\subsubsection{Tools.}

We selected three SBOM generation tools:
\\\syft\texttt{~v1.38.2}~\cite{syft},
\trivy\texttt{~v0.68.2}~\cite{trivy}, and
\texttt{\cdxgen~v12.0.0}~\cite{cdxgen}.

These versions were pinned to ensure experimental reproducibility.
\syft and \trivy were selected due to their frequent use in recent
empirical SBOM studies, their support for the ecosystems considered in
this work, and their adoption in DevSecOps workflows. \cdxgen was
included as a representative SBOM generator from the CycloneDX
ecosystem.  All three tools are configured to produce output in
CycloneDX~1.6 JSON format.

\smallskip

For lockfile-based campaigns, \cdxgen is invoked with the
\pkg{--no-install-deps} option to prevent additional dependency
resolution and ensure that generated SBOMs reflect only the
information available from the frozen lockfile. Although the presence
of a frozen lockfile should normally prevent dependency
reconstruction, this option provides an additional safeguard against
implicit installation or resolution steps performed by the tool. This
flag is disabled in the manifest-only setting
(Section~\ref{sec:manifest-only}), where no lockfile is available and
\cdxgen’s dependency resolution capabilities are explicitly evaluated.

\subsubsection{Campaigns Design.}

For each ecosystem (\js and \rust), we define three campaigns varying
the dependency information available to the tools at generation time,
for a total of six campaigns.

The first campaign assesses tool behavior in the manifest-only setting
(Section~\ref{sec:manifest-only}), when only manifest files are
available. 

The remaining two campaigns are lockfile-based. They rely on the same
frozen dataset (Section~\ref{sec:dataset}) and differ only in the
lockfiles made available to the tools during SBOM generation. The
\emph{clean\_root} method exposes only the root lockfile, making it
the only available lockfile reference, while temporarily removing
lockfiles located in subdirectories. Conversely, the
\emph{clean\_deep} method exposes all lockfiles present in the frozen
snapshot, including both root and subdirectory lockfiles.

These two campaigns evaluate how SBOM generators behave under
different levels of dependency visibility and lockfile completeness.

\subsection{Detection and Normalization Pipeline}
\label{sec:pipeline}

A naive comparison between SBOM output and lockfile content counts many apparent mismatches that are not actual errors, since the same dependency can be represented differently depending on the tool. 
Before computing any metric, we apply a normalisation pipeline that aligns these representations so that identical dependencies are recognised as such by our experimental framework, regardless of how each tool expresses them. 
This pipeline operates only on the data extracted for comparison, never on the SBOM artefacts themselves, and no information is discarded: each normalisation is recorded, which makes it possible to trace what diverges and why. 
Beyond alignment, the pipeline also allows identifying the nature of differences that persist after normalisation: a \emph{representation difference}, where both sides refer to the same dependency but express it in distinct forms, a \emph{design choice}, where the tool makes a deliberate decision that diverges from the baseline, or a \emph{tool error}, where the tool output is simply incorrect, reporting  a component that cannot be traced back to any real dependency.

\subsection{Metrics}
We measure three quantities for each tool and campaign.

\textbf{Coverage} is the fraction of lockfile packages reported by the tool,
computed separately for the full dependency set, production dependencies only,
and development dependencies only. It corresponds to recall against the
lockfile baseline, and is reported as a per-project mean across all
contributing projects in a campaign.

\textbf{Extra packages} are packages reported by the tool but absent from
the lockfile. They are counted per project and per tool, and examined
qualitatively to identify their cause.

\begin{sloppypar}

\textbf{Divergences} are the points where tools disagree with one another, as identified by the normalisation pipeline. They are recorded and quantified, providing a data source on the precise packages where tools differ and on the reasons behind these discrepancies.

All metrics are computed after applying the normalisation pipeline described in Section~\ref{sec:pipeline}.

\end{sloppypar}


\section{Results}
\label{sec:results}
This section presents the results of our empirical evaluation. We first report coverage across tools and ecosystems (RQ1), then analyze the sources of divergence between tool output and the lockfile baseline (RQ2) and finally assess how completely tools populate expected fields for the components they do report (RQ3).

 \begin{figure}[t]
    \centering
    \includegraphics[width=\columnwidth]{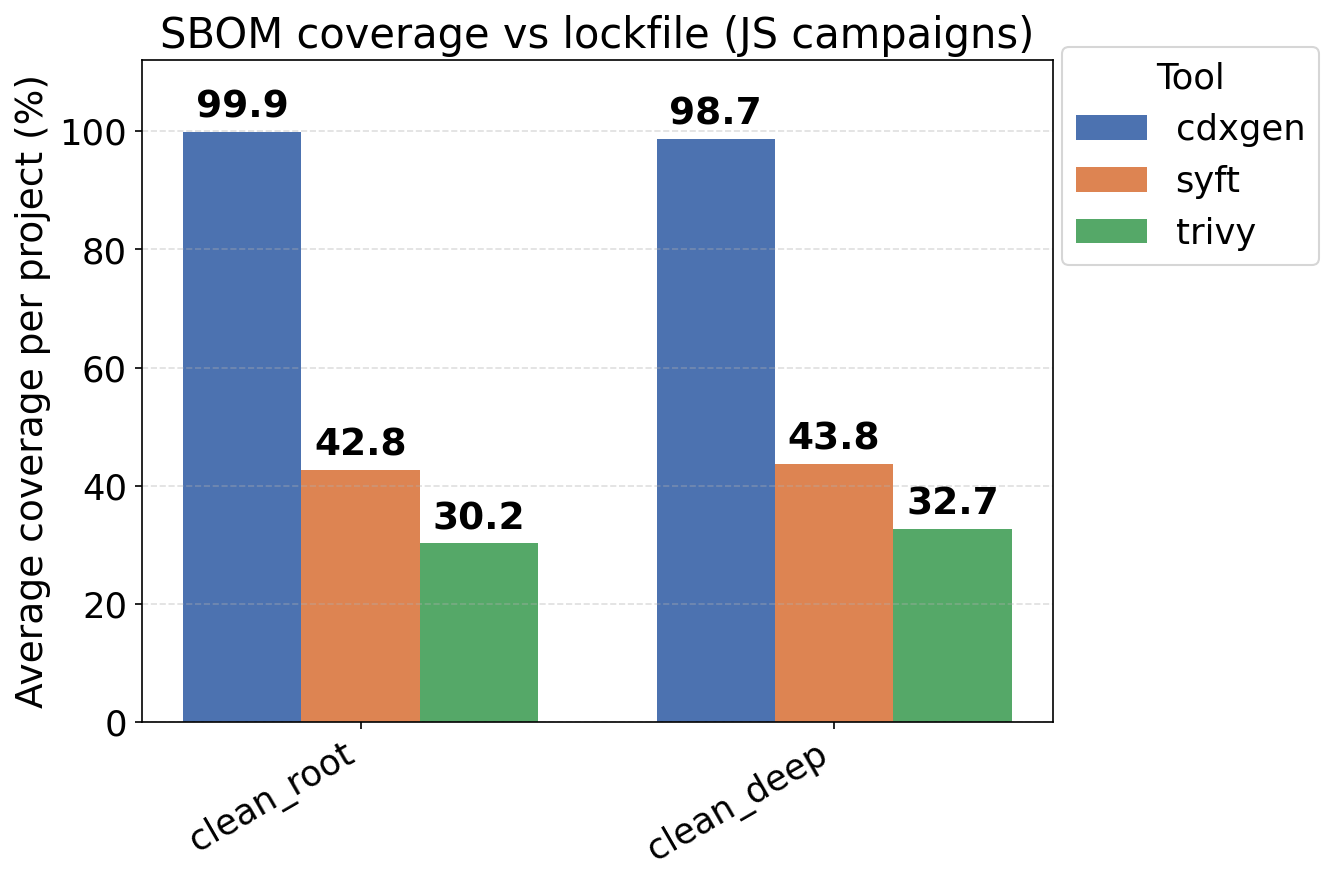}
    \caption{Average SBOM coverage against the lockfile baseline, JS corpus.}
    \label{fig:coverage-js}
  \end{figure}

  \begin{figure}[t]
    \centering
    \includegraphics[width=\columnwidth]{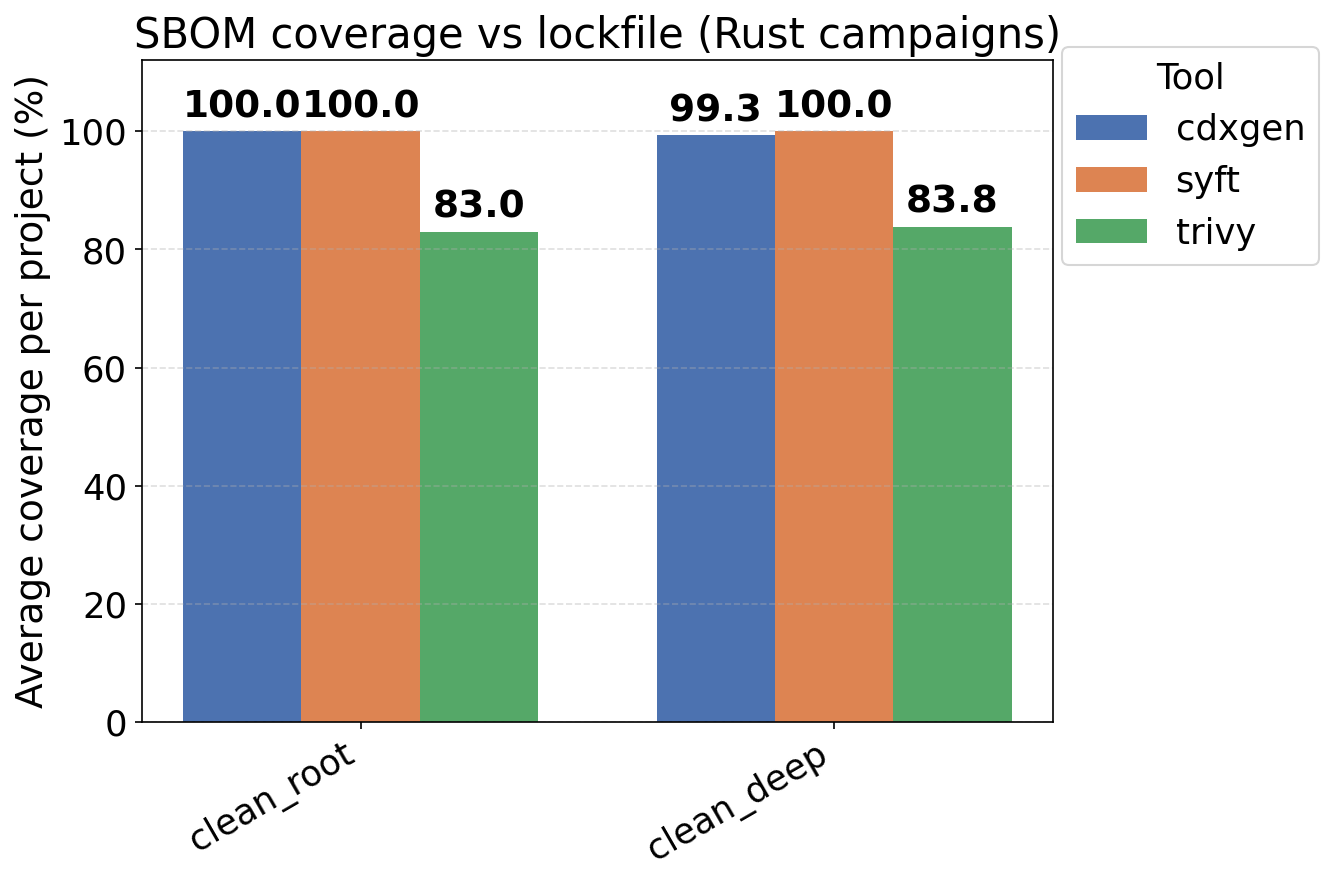}
    \caption{Average SBOM coverage against the lockfile baseline, \rust corpus.}

    \label{fig:coverage-rust}
  \end{figure}


\subsection{RQ1: Coverage Across Tools and Ecosystems and variation across dependency scopes}
\label{sec:rq1-results}

We address this research question through the \textbf{two lockfile-based campaigns} introduced in Section~\ref{sec:methodology}, \emph{clean\_root} and \emph{clean\_deep}, which share the same frozen corpus and differ only in the scope of lockfiles exposed to the tools at SBOM generation time. Figures~\ref{fig:coverage-js} and~\ref{fig:coverage-rust} report the average coverage achieved by each tool under these two campaigns, for \js and \rust respectively, computed after applying the normalization pipeline (Section~\ref{sec:pipeline}).

The tool \cdxgen  performs consistently well across both ecosystems and both installation scopes, reaching \covCdxgenJSd on \js and \covCdxgenRd on \rust under \emph{clean\_deep} (\covCdxgenJSr and \covCdxgenRr under \emph{clean\_root}), though not quite reaching 100\,\%, a residual gap we trace below. \syft and \trivy both exhibit a sharp ecosystem-dependent inversion, though with different magnitudes. Under \emph{clean\_deep}, \syft reports only \covSyftJSd of lockfile packages on \js projects, yet reaches \covSyftRd on \rust projects (\covSyftJSr and \covSyftRr under \emph{clean\_root}). \trivy follows a similar pattern, covering \covTrivyJSd on \js and \covTrivyRd on \rust under \emph{clean\_deep} (\covTrivyJSr and \covTrivyRr under \emph{clean\_root}). 

\medskip

 We investigated \cdxgen's residual gap and traced it to a hardcoded directory exclusion list in \cdxgen
  12.0.0. This list includes \path{/examples/}, which causes any lockfile located under a directory named exactly \texttt{examples} to be silently skipped, a simple string-matching rule, not a structural limitation of \cdxgen's dependency resolution. Its effect varies by setting:

  \begin{sloppypar}
  
  \begin{itemize}
      \item \rust: this rule, together with a few related exclusions (\path{docs/},
  \path{tests/}, \path{.github/}), explains 100\,\% of \cdxgen's missing packages under \emph{clean\_deep}, affecting a small number of multi-crate projects (\path{sea-orm},
  \path{cargo-leptos}, \path{trunk}).
      \item \js: the rule explains some, but not all, of the lockfiles \cdxgen never captured under
  \emph{clean\_deep}.
      \item \emph{clean\_root}: the rule never triggers, since exposing only the root lockfile means no
  lockfile can ever sit under an \path{examples/} subdirectory, which is why \cdxgen's scores are
  higher there.
  \end{itemize}

  \end{sloppypar}

In the \textbf{manifest-only setting}, that is, when projects are submitted without any lockfile, we observe two different behaviors across languages. For \js, \cdxgen attempts to install dependencies to generate a lockfile and report the full dependency graph; when the installation succeeds, the dependencies reported in its SBOM closely match those produced by our own installation, though we observe minor gaps for a small number of projects. 
For \rust, \cdxgen behaves differently: it performs no installation and only reads the manifest, so it naturally cannot report transitive dependencies. Moreover, even within the manifest, it does not report all declared dependencies, notably missing those declared as dev or build dependencies. Furthermore, the reported versions are not pinned, they are the version specifiers from the manifest rather than the resolved versions from a lockfile, meaning they do not match our lockfile-based baseline. Compared to this baseline, \cdxgen therefore only reaches a coverage of 3.2\%. For \trivy and \syft, the behavior is the same across both ecosystems: without a lockfile, both tools produce an empty SBOM, as they require a resolved lockfile to enumerate components and do not perform automatic installation.

\medskip 

Overall, these results confirm that the presence of a lockfile is a key determinant of SBOM completeness. However, even in the frozen setting, coverage remains imperfect for some tools, and its absence further degrades results, leading to partial (\cdxgen) or empty (\trivy, \syft) SBOMs in the manifest-only setting.


\subsection{RQ2: Mechanisms of Divergence}
\label{sec:rq2-results}
\subsubsection{Scope divergence: devDependencies handling}

\syft and \trivy both exhibit an abnormally low \js coverage, as both exclude devDependencies from their SBOM, but not in the same way. In \syft, the exclusion is a design choice applied inconsistently depending on the package manager used, while in \trivy, the exclusion is enabled by default but an implementation error causes it to partially fail, letting some transitive dependencies through.

\smallskip

\syft's inconsistency is package-manager-dependent: in the \npm project \pkg{js-framework-benchmark}, it reports 100\% of production dependencies (2249/2249) but excludes devDependencies entirely (0/4930). In the \yarn project \pkg{chakra-ui-vue} and the \pnpm project \pkg{ember.js}, by contrast, it reports both dependency types at 100\% (3164/3164 and 28/28 for \pkg{chakra-ui-vue}; 372/372 and 1724/1724 for \pkg{ember.js}), applying no distinction at all between production and development dependencies for these two formats. The same inconsistency appears in \rust: \syft reports devDependencies present in \texttt{Cargo.lock} despite applying exclusion logic elsewhere. This is not a bug in the traditional sense, but a design choice that has never been applied uniformly across the formats \syft supports.

\medskip

\begin{figure}[ht]
\centering
{\small\textbf{Expected structure (lockfile)}}\\[1mm]
{\scriptsize\textit{the entire subtree should be excluded from the SBOM}}\\
\vspace{2mm}
\includegraphics{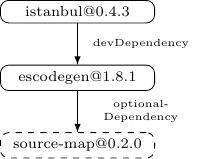}
\vspace{6mm}\\
{\small\textbf{What \trivy produces (SBOM)}}\\[1mm]
{\scriptsize\textit{\texttt{source-map@0.2.0} attached directly to \texttt{root package}}}
\vspace{2mm}
\includegraphics{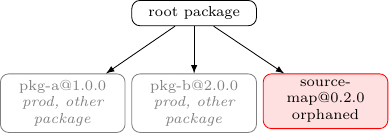}
\caption{Orphaned node produced by \trivy on a \yarn lockfile project :
the excluded dependency's child is incorrectly attached directly to the root package.}
\label{fig:trivy-yarn-orphan}
\end{figure}

\begin{sloppypar}

\trivy's divergence is of a different nature: the tool excludes development dependencies by default, but the exclusion logic is faulty. This malfunction manifests in \yarn lock files, as illustrated by the \pkg{Google Play Music Desktop Player} project. In the lockfile, \pkg{escodegen@1.8.1} is a direct development dependency (parent: \pkg{istanbul@0.4.3}), and one of its dependencies, \texttt{source-map@0.2.0}, is declared as an \pkg{optionalDependency} of \pkg{escodegen}, not as a devDependency itself, but it inherits its parent's dev status since \pkg{escodegen@1.8.1} is not reachable through any other path. Neither should therefore appear in the SBOM. \trivy does exclude \pkg{escodegen@1.8.1}, but not \pkg{source-map@0.2.0}, which ends up attached directly as a root node in the dependencies section of its SBOM, as though it were a top-level dependency, rather than being excluded along with its parent. The exclusion thus stops at the first level (the package directly marked as dev), without propagating to its transitive dependencies, as shown in Figure~\ref{fig:trivy-yarn-orphan}.

\end{sloppypar}


\subsubsection{Over-reporting mechanisms relative to the lockfile-based baseline}

Beyond exclusion mechanisms, \cdxgen and \trivy also report components that fall outside our lockfile-based baseline by design, through three distinct detection mechanisms summarized in Table~\ref{tab:overreporting}.

 \textbf{Filename-based detection.} \cdxgen's filename-based detection heuristic scans all files in the project independently of any lockfile or manifest, applying a regular expression to filenames to identify a <name>-<version> pattern. The resulting component is tagged with the filename detection technique and the lowest confidence score on \cdxgen's scale (0.25, versus 1.0 for a lockfile-derived entry or 0.7 for a manifest entry), explicitly signaling the absence of any verification against a registry or declared entry. This mechanism specifically targets vendored code, libraries manually copied into the project without going through a package manager. The heuristic is not perfect: in some cases, it incorrectly detects version-like elements embedded in source code comments, producing invalid entries such as \pkg{for@license}, which cannot be mistaken for real dependencies but still increase the component count.

 \smallskip

\begin{sloppypar}
\textbf{bower.json parsing.} \pkg{bower.json} is an older \js manifest format used for frontend dependencies before the widespread adoption of \npm. \cdxgen actively parses \pkg{bower.json} as a fully-fledged declaration source, on par with \pkg{package.json}, and extracts the declared name and version with the same reliability. This is therefore not a detection reliability issue, but a scope issue: \pkg{bower.json} predates modern lockfiles and was never part of our baseline, which is limited to lockfile formats.
\end{sloppypar}

\smallskip

\textbf{bun.lock parsing.} The bun.lock lockfile format, used by the Bun package manager, constitutes a third source of divergence. Some corpus projects include a bun.lock alongside an \npm lockfile; \trivy detects and parses it directly, tagging the resulting components as bun-typed, while \syft and \cdxgen do not read it. Since bun.lock is not part of our baseline, only bun-typed components that do not already appear in a recognized lockfile are counted.

\begin{table}[t]
    \centering
    \small
    \caption{Over-reporting mechanisms relative to the lockfile-baseline reference (\js).}
    \label{tab:overreporting}
    \resizebox{\columnwidth}{!}{%
    \begin{tabular}{lllr}
    \toprule
    \textbf{Mechanism} & \textbf{Tool} & \textbf{Example} & \textbf{Total} \\
    \midrule
    Filename-based detection & \cdxgen & jquery@3.2.1 (\pkg{fast-xml-parser}) & 786 \\
    bower.json parsing & \cdxgen & alt@0.18.3 (\pkg{alt}) & 650 \\
    bun.lock parsing & \trivy & \pkg{react-native-root-toast} & 557 \\
    \bottomrule
    \end{tabular}%
    }
  \end{table}


\subsubsection{Identifier representation divergence}

Unlike the mechanisms above, the following phenomena do not reflect errors relative to the baseline, but representational choices that complicate cross-tool matching: the same real-world component may be reported under different identifiers depending on the tool. Table~\ref{tab:identifier-divergence} summarizes these mechanisms.

\smallskip
\begin{sloppypar}

\textbf{Package aliasing.} Some lockfile entries use a key of the form \pkg{alias@npm:real-package@version}, where the alias and the real registry name coexist in the same entry. For example, in the \pkg{spectrum} project, the alias \pkg{react-dom} points to the real package \pkg{@hot-loader/react-dom@17.0.2}. When tools differ in the name they report, the resulting PURLs diverge, causing the same component to appear as two distinct entries across SBOMs. In the example above, \cdxgen and \trivy report the alias \pkg{react-dom}, while \syft reports the real scoped name \pkg{@hot-loader/react-dom}. Our pipeline normalizes all reported names to the real \npm package name extracted from the lockfile. 

\end{sloppypar}

\smallskip

\begin{sloppypar}
\textbf{Peer-dependency suffixes.} In \pnpm v5/v6 lockfiles, the snapshot key encodes peer dependency context directly into the version string, for example, in the \pkg{eslint-plugin-prettier} project, \pkg{@graphql-tools/wrap@11.1.16_graphql@16.14.2}, where \pkg{graphql@16.14.2} is the resolved peer dependency context, not part of the actual version. \cdxgen and \syft report the raw suffixed version instead of \pkg{@graphql-tools/wrap@11.1.16}, producing a different PURL, whereas \trivy handles this case correctly.
\end{sloppypar}

\smallskip

\textbf{Symlink resolution.} In some lockfiles, the same package appears under two distinct entries: a symlink entry (marked link: true) pointing to the source, and an actual source entry carrying the version. \syft indexes both entries, which produces an additional UNKNOWN component in the SBOM for the symlink entry, since it carries no version. In the \pkg{robot} project, the lockfile entry \pkg{node_modules/lit-robot} is a link to \pkg{packages/lit-robot}, which carries the real version \texttt{3.0.0}: \syft's SBOM contains both \pkg{lit-robot@UNKNOWN} (symlink entry) and \pkg{packages/lit-robot@3.0.0} (actual source entry). This mechanism is \syft-specific.

\smallskip

\textbf{Monorepo and vendor path prefixes.} \syft reports packages with their full path (packages/X, vendor/X) whereas the   other tools only report the basename, leading to divergent PURLs for the same component. This is precisely what happens to the resolved entry above, \syft reports \pkg{packages/lit-robot} instead of   \pkg{lit-robot} even once the version is correctly identified. A second example, in project \pkg{shields}, shows \syft reporting \pkg{vendor/http-deceiver} instead of   \pkg{http-deceiver}.

\smallskip

\begin{sloppypar}
\textbf{Hash, URL, and semver mismatches.} When a dependency is pinned to a git or tarball reference rather than a standard registry release, a tool may report that raw reference instead of the actual semantic version recorded in the same lockfile entry. For example, in the \pkg{v2.cn.vuejs.org} project, the dependency \pkg{hexo-generator-alias} is pinned to a GitHub fork at commit \texttt{67adb814...}, with real version \texttt{0.1.3}: \cdxgen reports the commit hash (\pkg{hexo-generator-alias@67adb814a76750f3c841825f955bd5dd92cd1f20}), \syft reports the full tarball URL (\url{https://codeload.github.com/chrisvfritz/vuejs.org-hexo-generator-alias/tar.gz/67adb814...}), and \trivy reports the semantic version \pkg{hexo-generator-alias@0.1.3}.
\end{sloppypar}

\smallskip

\textbf{\yarn workspace placeholders.} \syft relies exclusively on the lockfile: on \yarn Classic, which never records local workspace members in \texttt{yarn.lock}, the member is simply absent from the SBOM; on \yarn Berry, which does write a 0.0.0-use.local stub, \syft reports that placeholder verbatim. By contrast, cdxgen actively resolves the real version from the member's own \pkg{package.json}, regardless of whether \yarn Classic leaves no trace or \yarn Berry writes a placeholder, consistently across the dataset.

\begin{table}[t]
    \centering
    \small
    \caption{Identifier representation divergences (\js).}
    \label{tab:identifier-divergence}
    \resizebox{\columnwidth}{!}{%
    \begin{tabular}{lrrrr}
    \toprule
    \textbf{Mechanism} & \textbf{cdxgen} & \textbf{trivy} & \textbf{syft} & \textbf{Total} \\
    \midrule
    Package aliasing & 412 & 104 & 64 & 580 \\
    \pnpm peer-suffix & 104 & 0 & 104 & 208 \\
    Symlink resolution & --- & --- & 755 & 755 \\
    Path prefixing & --- & --- & 305 & 305 \\
    \bottomrule
    \end{tabular}%
    }
\end{table}


\subsubsection{Structural divergence: root component and dependency graph topology}

Beyond individual component discrepancies, tools also diverge in how they represent the root package and structure the overall dependency graph, a phenomenon observed in both \js and \rust.

Root package handling differs across tools. The tool \cdxgen includes the root package in \texttt{metadata.component}. \trivy only mentions the project path in \texttt{metadata.component}, similar to \syft. \trivy also adds a dependency that is simply the lockfile itself.

\smallskip

\begin{figure*}[t]
  \centering
  \includegraphics[width=0.85\textwidth]{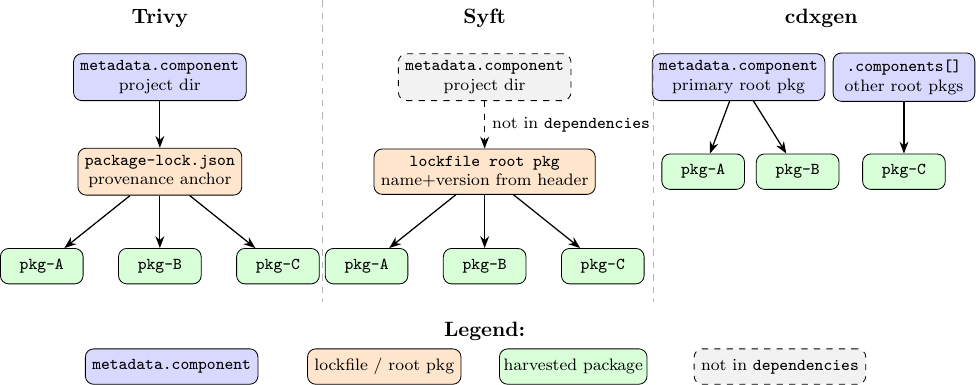}
  \caption{Dependency graph structure in the \texttt{dependencies} section across Trivy, Syft, and cdxgen (\js).}
  \label{fig:dep-graph-tools}
  \end{figure*}

\begin{sloppypar}
In \js, the three tools produce three distinct topologies, illustrated in Figure~\ref{fig:dep-graph-tools}. \trivy follows a three-level hierarchy: the scanned project directory is the top-level \pkg{metadata.component}, which depends on one or more lockfile components (e.g., \pkg{package-lock.json}) acting purely as provenance anchors, each in turn depending on the harvested packages. \syft also declares the project directory as \pkg{metadata.component}, but, unlike \trivy, never surfaces it as a root node in the dependencies section; instead, it introduces a separate lockfile-root component, taken from the lockfile's own header, which anchors the harvested packages. \cdxgen declares every lockfile's root package under \pkg{metadata.component} (nesting secondary ones under \pkg{metadata.component.components}), and uses their \pkg{bom-ref} values directly as   root nodes in the dependencies section.   
\end{sloppypar}

\smallskip

In \rust, the three topologies converge more closely: all three tools surface the root package in the components section, but \trivy and \cdxgen each add one extra node relative to \syft. \trivy turns the root crate into a genuine intermediate node rather than a mere provenance anchor, producing a four-level hierarchy (project directory, \texttt{Cargo.lock}, root crate, dependencies) instead of three. \cdxgen lists the root crate twice, once as \texttt{metadata.component} and once as a components-section entry sharing the same \texttt{bom-ref}.   \syft's behavior is unchanged from \js.


\subsection{RQ3: Spec Completeness}
\label{sec:rq3-results}

RQ1 and RQ2 ask whether tools capture the right components. This section asks a different question, evaluated directly on the produced SBOMs rather than on the subset of components that match the lockfile baseline reference: does the SBOM populate the fields that a specification-compliant consumer would be entitled to expect? 

\smallskip

We measure this against the NTIA's minimum elements (Supplier, Component Name, Version, Other Unique Identifiers, Dependency Relationship, Author of SBOM Data, Timestamp), together with three additional CycloneDX fields that vary widely across tools in practice: licenses, hashes, and external references. CISA's 2025 update to the minimum elements, a public comment draft at the time of writing~\cite{cisa2025minimum}, notably promotes \textit{License} and \textit{Component Hash} to formal minimum elements in their own right~\cite{cisa2025minimum}, which reinforces the relevance of examining these two fields alongside the original NTIA set. Tables~\ref{tab:spec-completeness-js} and~\ref{tab:spec-completeness-rust} report these rates under \emph{clean\_deep} for \js and \rust respectively. Component name, version, unique identifiers (PURL or CPE), dependency relationships, and timestamp are populated by all three tools on both ecosystems at 99.7\,\% or above, we therefore focus below on the fields that diverge sharply.

\smallskip

\begin{table}[ht]   
  \centering
  \small
  \caption{Spec completeness (JS, js\_clean\_deep)}
  \label{tab:spec-completeness-js}
  \resizebox{\columnwidth}{!}{%
  \begin{tabular}{lrrr}
  \toprule
  Field & cdxgen & Syft & Trivy \\ 
  \midrule
  Component name  & 100.0\% & 100.0\% & 100.0\% \\
  Version  & 100.0\% & 100.0\% & 100.0\% \\
  PURL  & 100.0\% & 100.0\% & 100.0\% \\
  CPE  & 0.0\% & 100.0\% & 0.0\% \\
  Supplier  & 0.0\% & 0.0\% & 0.0\% \\
  Licenses  & 75.8\% & 60.2\% & 0.0\% \\
  Hashes  & 98.2\% & 0.0\% & 0.0\% \\
  External references  & 0.0\% & 0.0\% & 0.0\% \\
  Dependency relationships  & 100.0\% & 99.7\% & 100.0\% \\
  Timestamp  & 100.0\% & 100.0\% & 100.0\% \\
  Author of SBOM data, strict  & 100.0\% & 0.0\% & 0.0\% \\
  Tool Name  & 100.0\% & 100.0\% & 100.0\% \\
  Generation context  & 100.0\% & 0.0\% & 0.0\% \\
  \bottomrule
  \end{tabular}%
  }
\end{table}
\begin{table}[ht]
  \centering
  \small    
  \caption{Spec completeness (Rust, rust\_clean\_deep)}
  \label{tab:spec-completeness-rust}
  \resizebox{\columnwidth}{!}{%
  \begin{tabular}{lrrr}
  \toprule  
  Field & cdxgen & Syft & Trivy \\
  \midrule
  Component name  & 100.0\% & 100.0\% & 100.0\% \\
  Version  & 100.0\% & 100.0\% & 100.0\% \\
  PURL  & 100.0\% & 100.0\% & 100.0\% \\
  CPE  & 0.0\% & 99.9\% & 0.0\% \\
  Supplier  & 0.0\% & 0.0\% & 0.0\% \\
  Licenses  & 0.0\% & 0.0\% & 0.0\% \\
  Hashes  & 97.8\% & 0.0\% & 0.0\% \\
  External references  & 0.0\% & 97.5\% & 0.0\% \\
  Dependency relationships  & 100.0\% & 100.0\% & 100.0\% \\
  Timestamp  & 100.0\% & 100.0\% & 100.0\% \\
  Author of SBOM data, strict  & 100.0\% & 0.0\% & 0.0\% \\
  Tool Name  & 100.0\% & 100.0\% & 100.0\% \\
  Generation context  & 100.0\% & 0.0\% & 0.0\% \\
  \bottomrule
  \end{tabular}%
  }
  \end{table}

The clearest gap concerns the \textbf{Supplier field, populated by none of the three tools on either ecosystem} (Tables~\ref{tab:spec-completeness-js}--\ref{tab:spec-completeness-rust}); CISA's 2025 draft renames this element \textit{Software Producer} and sharpens its definition to the entity that produces the software itself, as opposed to a distributor~\cite{cisa2025minimum}, making the gap even harder to fill from SBOM metadata alone. The remaining divergent fields show no single pattern common to all four: the CPE component of unique identifiers and hashes split along tool lines, each populated almost exclusively by one tool, though not the same tool for both; licenses split instead along ecosystem lines; and external references vary by both tool and ecosystem simultaneously (Tables~\ref{tab:spec-completeness-js} and~\ref{tab:spec-completeness-rust}). 

\smallskip

The Author of SBOM Data field, in the strict NTIA sense, is populated only by \cdxgen; \syft and \trivy leave it empty, although the identity of the generating tool can still be recovered through other SBOM metadata in all three cases (a tool-fallback measure we compute separately, at 100\,\% for all three), a distinction CISA's 2025 draft now formalizes as two separate elements, \textit{SBOM Author} and \textit{Tool Name}~\cite{cisa2025minimum}, the latter matching exactly what our tool-fallback measure captures. CISA's 2025 draft also introduces \textit{Generation Context} as a new minimum element, recording the software lifecycle phase (before, during, or after build) at which the SBOM was generated~\cite{cisa2025minimum}; only \cdxgen populates it (always tagged pre-build), leaving consumers of \syft or \trivy SBOMs with no way to tell, from the SBOM alone, what kind of artifact was analyzed.


\section{Discussion} 
\label{sec:discussion}

Our results show that the divergences documented in prior work are, for the most part, neither random nor mysterious: they trace back to identifiable design choices and representation differences (RQ1, RQ2), rather than to chance. RQ3 reveals a separate, independent problem: the fields a spec-compliant SBOM consumer would expect are \textbf{populated inconsistently across tools}, and for at least one mandatory field, never at all. Solving the coverage   and representation problems would not, by itself, be enough to produce a compliant SBOM.

\subsection{SBOM and tools limitations}
\label{sec:sbom-tools-limit}

Stepping back, the coverage and mechanism divergences documented in RQ1 and RQ2 share a common root cause: the demand placed on SBOM generation tools is not precise enough. Current SBOM standards, including CycloneDX and SPDX, provide detailed guidance on how a dependency should be represented once selected, but do not precisely define \emph{which dependencies} should be included in the first place. To improve interoperability and enable reproducible comparisons between SBOM generation tools, we believe that several aspects of SBOM generation deserve further standardization.

\subsubsection{Scope of dependencies.} A source of divergence concerns \emph{which} dependency types are in scope (production, development, test, documentation, optional, peer, or workspace), a divergence we document directly in our devDependencies findings (Section~\ref{sec:rq2-results}). \citet{Tobar2025} and \citet{rabbi2025claim} present build-time, testing, or environment-specific dependencies as a source of false positives, arguing that they contaminate the generated SBOM with components that are not part of the \emph{deployed} application. 
In our work, we deliberately do \emph{not} adopt this filtering perspective: we consider build, test, development, and runtime dependencies as all valuable when generating SBOMs, and argue that both a \emph{development SBOM} and a \emph{deployment SBOM} are relevant, complementary artifacts, capturing two distinct moments of the software lifecycle. The former exposes the full dependency surface accessible to contributors and CI/CD pipelines, relevant when reasoning about supply-chain attacks targeting build or test tooling, while the latter reflects the actual attack surface of the application as it runs in production. Consequently, what is considered a "false positive" in the context of a \emph{deployment} SBOM may constitute legitimate, security-relevant information in the context of a \emph{development} SBOM.
Indeed, development dependencies remain critical for assessing supply-chain risk and have themselves been the target of supply-chain attacks~\cite{duan2020towards,ohm2020backstabbers,zahan2021weaklinks}. This is precisely why our lockfile baseline deliberately includes components originating from test directories and fixtures, as well as build-time and development dependencies, rather than being restricted to runtime/production-only components. Rather than leaving this choice to each tool discretion, SBOMs could record each component scope through a standardized attribute, making filtering an explicit, auditable step rather than a default behavior.


 \subsubsection{Origin of dependencies.} A source of divergence concerns how components are discovered. Our RQ2 results show tools relying on markedly different sources: lockfiles, manifests, filenames of vendored code, or alternative lockfile formats such as bower.json and bun.lock (Section~\ref{sec:rq2-results}). Given the diversity of programming languages and package management ecosystems, standardizing the discovery heuristics themselves is inherently complex; what can and should be standardized, however, is the way their outcome is recorded. We suggest that provenance tagging become a standardized, cross-tool attribute rather than one tool's internal convention. This would let consumers distinguish components that are explicitly declared from those merely inferred by heuristics of varying reliability.

\subsubsection{Dependency graph structure.} A source of divergence concerns the representation of the root component and the dependency graph itself. Our structural findings (Section~\ref{sec:rq2-results}) show the three tools adopt three distinct topologies for the same \js project, and a further split on \rust, which reduces interoperability with downstream SCA tooling and complicates cross-tool comparison. Consequently, we advocate for future revisions of SPDX and CycloneDX to define a canonical graph structure, specifying, at minimum, how the root component is identified.

\smallskip

Beyond these standardization gaps, our results also surface concrete near-term risks for CRA compliance. The Cyber Resilience Act (CRA) turns SBOM generation from a voluntary security practice into a compliance obligation, and in practice this obligation will overwhelmingly be discharged using widely available open-source generators such as the three studied here. Our RQ1 results show (Section~\ref{sec:rq1-results}) that this is not a neutral choice: depending on whether a \js publisher generates its SBOM with \syft, \trivy, or \cdxgen, \textbf{the resulting inventory's completeness varies substantially} depending on which tool a \js publisher uses. A publisher who follows the letter of the regulation by producing an SBOM has no straightforward way to know whether they have produced a complete one.

\subsubsection{Lockfile vs. manifest}
A more severe version of this problem occurs when a project has no lockfile at all, a common case for libraries, which often exclude \texttt{package-lock.json} or \texttt{Cargo.lock} from version control. In this manifest-only case, \trivy and \syft do not just under-report: \textbf{they produce an empty SBOM}, with no error or warning shown to the operator. \cdxgen avoids this problem for \js by trying to install the project and build a full dependency graph, though the install does not always succeed, but not for \rust, where it just reads the manifest file directly. This means it only reports unpinned, direct dependencies. A publisher whose repository policy leaves out lockfiles, whether on purpose or by mistake, would therefore get an SBOM that looks plausible but is actually empty or badly incomplete from two of the three tools we test, and a partial one from the third, depending entirely on which 
ecosystem the project uses.


\subsubsection{Dependency identifiers}
A second, independent exposure arises even when a dependency is correctly detected. The identifier-representation divergences, an \npm alias reported under its alias rather than its real registry name, a \pnpm peer-dependency suffix left in the version string, the same lockfile entry reported at times as a git commit hash, at times as a URL, at times as a semantic version, result in different PURLs being generated. Since vulnerability analysis tools rely on these identifiers to correlate software components with known vulnerabilities, discrepancies in generated PURLs may lead to different analysis outcomes depending on the identifier being used. Such discrepancies could potentially undermine the effectiveness of the primary use case (vulnerability management) that motivates the adoption of SBOMs under the CRA. 
Under this reading, \textbf{tool choice is no longer just an engineering decision, it is a compliance-exposure question} that publishers, or anyone generating an SBOM, currently have no practical way to evaluate. Our work provides a first step towards a method and tool to address this issue.

\subsubsection{SBOM fields}
A third, distinct gap concerns the SBOM's content itself, independent of which components it lists. The fields a specification-compliant SBOM consumer would expect are populated unevenly across tools (RQ3): the CPE component of unique identifiers and hashes split along tool lines, each populated almost exclusively by one tool though not the same tool for both, licenses split along ecosystem lines, and external references vary by both tool and ecosystem simultaneously. A publisher therefore does not obtain an SBOM carrying the same information regardless of which tool generated it. The clearest case is the Supplier field: none of the three tools populate it, one of NTIA's original minimum elements, sharpened in CISA's 2025 draft update to "Software Producer"~\cite{cisa2025minimum}, on either ecosystem. A publisher relying on any of these tools cannot meet this transparency obligation through automated SBOM generation alone.

\subsection{Limitations to our work} \label{sec:limitations}

\subsubsection{Dataset selection.} Our dataset is limited to GitHub repositories with at least 1,000 stars, favoring actively  maintained, popular projects over the smaller, less curated ones that make up most real-world dependencies. We adopted this threshold as a practical proxy for project maturity: highly-starred repositories are more likely to be well-maintained projects with a \pkg{Cargo.toml} or \pkg{package.json} manifest at the root, which is required for our SBOM generation tools to work (on \rust and \js projects respectively). We do not measure how this affects our coverage figures, and cannot rule out that they would differ on a less popularity-skewed sample.

\subsubsection{Durability of tool-specific findings.} Our results are tied to specific tool versions (\syft~1.38.2,   \trivy~0.68.2, \cdxgen~12.0.0), and these tools evolve quickly enough that their behavior can change substantially   in a short amount of time. \citet{Zhou_2026}, for instance, validated lockfiles as ground truth for Python,   \rust, Ruby, and PHP but excluded \js after finding that the contemporary tool versions they used (\trivy~0.66.0,   \syft~1.33.0) produced empty SBOMs for \texttt{package-lock.json}. Using newer versions (\trivy~0.68.2,   \syft~1.38.2), we find that neither tool produces empty SBOMs anymore, but both now produce partial SBOMs instead.   The contrast with \citet{rabbi2024sbom} and \citet{rabbi2025claim} points to the same pattern.

Conclusions drawn here may thus have a limited shelf life and would benefit from periodic re-evaluation as tools are updated. However, such conclusions can clearly be considered more as fuel for test cases, to assess the quality of upcoming tools and versions.

\subsubsection{Field presence versus field correctness.} RQ3 measures whether a given specification field \textbf{is populated, not whether its content is correct}. A component whose \texttt{licenses} field is non-empty is counted as compliant in our measurement, even if the reported license is incorrect; the same applies to hashes, external references, and the other fields in Tables~\ref{tab:spec-completeness-js} and~\ref{tab:spec-completeness-rust}.

\section{Conclusion} \label{sec:conclusion}

 We generated SBOMs with three widely used tools (\syft, \trivy, and \cdxgen) over the same corpus of \js and
  \rust projects, and compared them against a lockfile-based reference. 
  
Our key findings are as follows:

\begin{itemize}

\item \textbf{Tool coverage diverges substantially (RQ1).} The tools disagree substantially on which
dependencies they report, and this disagreement worsens, or turns into an empty SBOM, when a project has no
lockfile.

\item \textbf{Divergences are explainable, not random (RQ2).} A large share of the disagreement traces back to
specific, identifiable causes rather than chance: deliberate design choices such as excluding devDependencies
(a design choice), and representation artifacts such as aliasing, peer-suffixes, or path-prefixing that
change a component's identifier without changing the underlying dependency (a representation difference).

\item \textbf{Spec-field completeness is uneven (RQ3).} The fields a specification-compliant SBOM is expected
to carry are populated unevenly across tools, and at least one mandatory field, Supplier, is never populated at
all.

\item \textbf{Standardization gaps span scope, origin, and structure.} (Section~\ref{sec:sbom-tools-limit}). 
Current SBOM standards do not precisely
define which dependencies should be included, how their provenance should be recorded, or how the dependency
graph should be structured, leaving each choice to individual tools. We argue that scope and origin should each
be recorded through standardized, cross-tool attributes, and that SBOM formats should specify a canonical graph
structure, rather than leaving these choices to each tool's default behavior.

\item \textbf{Tool choice is a compliance risk.} SBOM completeness and content currently depend on which tool
is used, not only on the project being scanned, which undermines the transparency and vulnerability-management use
cases that motivate SBOM adoption in the first place: a scan run against a tool-reported identifier can silently
diverge from a scan run against the true package.
\end{itemize}

This does matter for the CRA: as it enters into force, publishers will rely on exactly these tools to produce their
mandated SBOMs, so \textbf{the tools themselves risk becoming a source of non-compliance that publishers have no way to detect}.

\smallskip
  
\noindent\textbf{Future work.} Two directions follow directly from this work. First, extending the comparison to
additional ecosystems and package managers (Python, Java, ...) would help establish whether the mechanisms we identify here,
design choices and representation differences, uneven field completeness, generalize beyond \js and \rust, or are
ecosystem-specific. Second, and more directly, the divergence cases we catalogued in this paper are not only findings: they are also 
subtle, controlled, concrete test cases. We plan to turn this catalogue into a benchmark for SBOM generation tools, so that
a given tool or version can be checked against each identified edge case individually, rather than relying on
unstructured, large-scale comparisons like ours to surface regressions after the fact.

\bibliographystyle{ACM-Reference-Format}
\bibliography{references}

\end{document}